\documentclass[twocolumn]{aa}
\usepackage{astrobib}
\usepackage{journals}
\usepackage{graphicx}
\usepackage{txfonts}
\def\kms  {km~s$^{-1}$}
\def\water {H$_2$O}
\begin{document}
   \titlerunning{The nuclear region of NGC\,253}
   \authorrunning{Brunthaler et al.}
   \title{Evidence of a pure starburst nature of the nuclear region 
          of NGC\,253}


   \author{A. Brunthaler
          \inst{1}
          \and
	  P. Castangia\inst{1,2}
          \and
	  A. Tarchi\inst{2,3}
	  \and
          C. Henkel\inst{1}
	  \and
          M. J. Reid\inst{4}
          \and
          H. Falcke\inst{5,6}
          \and
          K. M. Menten\inst{1}
          }

   \institute{Max-Planck-Institut f\"ur Radioastronomie, Auf dem H\"ugel 69,
              53121 Bonn, Germany
              \and
	      INAF-Osservatorio Astronomico di Cagliari, Loc. Poggio dei Pini, Strada 54, 09012 Capoterra (CA), Italy
	      \and
	      INAF - Istituto di Radioastronomia, via Gobetti 101, 40129 Bologna, Italy
	      \and
              Harvard-Smithsonian Center for Astrophysics, 60 Garden Street,
              Cambridge, MA 02138, USA
              \and
               Department of Astrophysics, Radboud Universiteit
               Nijmegen, Postbus 9010, 6500 GL Nijmegen, The Netherlands
               \and
              ASTRON, Postbus 2, 7990 AA Dwingeloo, the Netherlands
            }

   \date{Received}

  \abstract{
We present high-resolution spectral line and continuum VLBI and VLA 
observations of the nuclear region of NGC\,253 at 22 GHz. While the water
 vapor masers in this region were 
detected on arcsecond and milliarcsecond scales, we could not detect any 
compact continuum emission with a 5$\sigma$ upper limit of $\sim$ 1 mJy. 
The observations reveal that the water maser emission is not related to 
a possible low-luminosity active galactic nucleus but is almost certainly 
associated with star-formation activity. Not detecting any compact 
continuum source on milliarcsecond scales also questions the presence of 
a -- previously assumed -- active nucleus in NGC\,253.
 
   {}

}
   \maketitle
%

\section{Introduction}

Extragalactic water masers in the 22.23508 GHz transition between the 6$_{16}$ 
and 5$_{23}$ rotational levels of ortho-H$_2$O are found in star-forming 
regions (e.g., M33: \citeNP{ChurchwellWitzelHuchtmeier1977}, LMC: 
\citeNP{ScaliseBraz1981}, IC\,10: \citeNP{BeckerHenkelWilson1993}; 
\citeNP{CastangiaTarchiHenkel2008}) and 
associated with an active galactic nucleus (AGN), either in accretion disks 
(e.g., NGC\,4258, \citeNP{MiyoshiMoranHerrnstein1995}; UGC\,3789, 
\citeNP{ReidBraatzCondon2008}), cloud-jet interactions 
(e.g., NGC\,1052, \citeNP{ClaussenDiamondBraatz1998}; NGC\,1068 
\citeNP{GallimoreHenkelBaum2001}; Mrk\,348, \citeNP{PeckHenkelUlvestad2003}), 
or nuclear outflows (Circinus, \citeNP{GreenhillBoothEllingsen2003}).
While the star forming masers are only detected in relatively nearby galaxies, 
the powerful AGN masers can be observed out to cosmological distances 
(\citeNP{BarvainisAntonucci2005}; \citeNP{ImpellizzeriMcKeanCastangia2008}). 
Extragalactic 
water masers are a 
powerful tool for investigating the nuclear environment of active galaxies and 
have also been used to measure the proper motions of members of the local group
of galaxies \cite{BrunthalerReidFalcke2005,BrunthalerReidFalcke2007}.

One of the nearest water masers outside the local group that can be used for
proper motion studies is located in NGC\,253, a prominent spiral galaxy in 
the nearby Sculptor group of galaxies. It has been known for more than 20 years
\cite{HoMartinHenkel1987} and is associated with the galaxy's mildly 
starbursting nuclear environment \cite{HenkelTarchiMenten2004}. NGC\,253 is 
commonly placed at a 
distance of 2.5 Mpc, but some studies favor smaller ($\le$2.3 Mpc: 
\citeNP{DavidgePritchet1990}) or larger (3.5 Mpc: 
\citeNP{RekolaRicherMcCall2005}; 3.3 Mpc: \citeNP{MouhcineFergusonRich2005}) 
distances. Throughout the paper we adopt the recent tip of the red giant 
branch distance estimate of 3.3 Mpc (1$''$ = 16 pc or 1 mas = 3300 AU). At 
this distance, one could measure the proper motion of NGC\,253 with an 
accuracy of $\sim$50 \kms\ within a few years.

Radio continuum images of NGC\,253 with arcsecond 
resolution show several compact radio sources  
(\citeNP{UlvestadAntonucci1997}, hereafter UA97). The strongest source (TH2 
in UA97) near the dynamical center of the galaxy is unresolved and may 
represent a low-luminosity active galactic nucleus (LLAGN) which is not 
dominating the nuclear environment. VLA A-configuration observations at 22 GHz 
(\citeNP{HofnerBaanTakano2006}, hereafter HBT06) have 
shown that the maser is not associated with TH2 but with 
another nearby continuum source (TH4 in UA97), which is probably a supernova 
remnant. 

Here we present deep 22 GHz VLBI and VLA observations with milliarcsecond 
(mas) and arcsecond resolution of the water maser and the continuum sources in
the central region of NGC\,253.

\section{Observations and Data reduction}
\subsection{VLBI observations}
We observed NGC\,253 with the NRAO\footnote{The National Radio Astronomy 
Observatory is a facility of the National Science Foundation operated under 
cooperative agreement by Associated Universities, Inc.} Very Long Baseline 
Array (VLBA) and the phased Very Large Array (VLA) in its C configuration
under program BB254 on 2008 March 24. We observed in two 16 MHz wide bands 
in dual circular polarization each. The 256 spectral channels in 
each band yielded a channel spacing of 62.5 kHz, equivalent to 0.84 
km s$^{-1}$, and covered a velocity range of 216 km s$^{-1}$. One frequency 
band was centered at an LSR velocity of 120 \kms (the peak of the maser 
emission), while the second band was centered at an LSR velocity of 
$-$95.72 \kms.

The observations involved rapid switching for 6 hours between the 
phase-calibrator VCS1~J0038$-$2459 from the VLBA Calibrator Survey 
\cite{BeasleyGordonPeck2002}, which is a compact background continuum 
source, and the target source NGC\,253. We switched sources every 50 seconds 
and achieved on-source integration times of $\sim$ 42 seconds for the VLBA 
antennas and $\sim$ 20 seconds for the slower VLA antennas. 
3C\,454.3 was also observed as fringe finder.

Before and after the phase referencing observation we included 
{\it geodetic-like} observations where we observed for 60 minutes $\sim$15 
strong radio sources ($>$ 200 mJy) with
accurate positions ($<$ 1 mas) at different elevations to estimate an
atmospheric zenith delay error in the VLBA calibrator model (see
\citeNP{ReidBrunthaler2004} and \citeNP{BrunthalerReidFalcke2005b} for a
discussion). Since the atmospheric zenith delay should be very similar for 
all VLA antennas, we used only a single VLA antenna in the {\it geodetic-like} 
observations. The other antennas were used to observe NGC\,253 in spectral 
line mode  (see the end of Sect.\,2.2).

The VLBI data were edited and calibrated using standard programs in the
Astronomical Image Processing System (AIPS). First, we applied the latest 
values of the Earth's orientation parameters. Zenith delay corrections 
were performed based on the results of the geodetic-like observations. 
A-priori amplitude calibration was applied using system temperature 
measurements and standard gain curves. We performed a ``manual 
phase-calibration'' using the data from 3C\,454.3 to remove instrumental 
phase offsets among the two frequency bands. Then, we fringe fitted the
data from VCS1~J0038$-$2459. Since the a priori amplitude calibration of the 
phased VLA was off by a large factor, we performed two steps of amplitude 
self-calibration on VCS1~J0038$-$2459. First, we changed only the gain of the 
VLA and included all antennas in the second step. Finally, we transfered 
the calibration to our target source NGC\,253.

\subsection{VLA observations}

NGC\,253 was observed with the VLA in its most extended A configuration
under program AC753 on 2004 November 2. We employed a single band of 12.5\,MHz 
subdivided in 64 channels of width 195.3\,kHz, yielding a channel spacing of 
2.63 \kms\, and covered a velocity range of 168 \kms. The observing band was 
centered at the frequency corresponding to the LSR velocity (120\,{\kms}) 
of the main maser component. The absolute amplitude calibration was obtained 
using 3C\,48 ($S_{\rm 22\,GHz}$=1.13\,Jy), while as phase calibrator we 
employed the compact source 0120$-$270 ($S_{\rm 22\,GHz}$=0.54\,Jy). The total 
observing time 
was 4.5 hours. We conducted pointing observations regularly and employed fast 
switching with 170\,s on the target source and 70\,s on the phase calibrator.

The data were calibrated in the standard manner using AIPS. The continuum 
emission
was subtracted from the spectral line data by fitting a straight line to the 
visibilities of the line free channels with the task UVLSF. This task also 
uses the fitted baseline to construct a line-free uv-dataset that was used 
to create a 
continuum map. Both line and continuum datasets were Fourier-transformed 
using natural weighting and then deconvolved utilizing the CLEAN algorithm.

The data from the phased VLA in our VLBI observation on 2008 March 24, 
obtained with the VLA in C configuration, were 
also correlated in {\it normal} VLA continuum mode with two frequency bands 
of 50 MHz width in dual polarization. The data was calibrated using standard 
techniques in AIPS. 3C\,48 was used for the flux density calibration. 
VCS1~J0038$-$2459 was then used as amplitude and phase calibrator before we 
performed two rounds of phase self-calibration and one phase \& amplitude 
self-calibration on NGC\,253. 

Since we used only a single VLA antenna during the {\it geodetic-like} 
observations, we could simultaneously observe NGC\,253 with the remaining 
antennas in 
spectral line mode. We used a total bandwidth of 12.5 MHz and 64 spectral 
channels, yielding a channel spacing of 2.63 \kms\, and covered a velocity 
range of 168 \kms. We employed fast switching between NGC\,253 and the nearby
phase calibrator 0120$-$270. The total integration time on NGC\,253 was 24 
minutes. The data was also calibrated using standard techniques in AIPS. 
3C\,48 was used for the flux density calibration. After amplitude and phase 
calibration on 0120$-$270 we performed first a phase self-calibration and then 
a phase \& amplitude self-calibration using our image from the continuum 
observation. Finally, we subtracted the continuum emission 
from the uv-data set.

\section{Results}
\subsection{VLBI Data}
The VLBI data were imaged with a Gaussian uv-taper of 350,000 k$\lambda$ 
and natural weighting to obtain the highest brightness sensitivity. We 
mapped two fields of 0.4$\times$0.4 arcsecond$^2$ centered at the positions 
of the VLA 
components TH2 and TH4. First, all channels in the line free frequency band 
were averaged to image the continuum emission of NGC\,253. The restoring 
beam was 2.44 $\times$ 1.11 mas$^2$ with a position angle of 12$^\circ$. 
No continuum emission was detected with a 5$\sigma$ upper limit of 1.1 mJy
for both fields.  Since no maser is known in the field of TH2, we imaged 
this field also with both frequency bands to obtain a deeper image. The 
restoring beam was 2.63 $\times$ 1.31 mas$^2$ with a position angle of 
14$^\circ$. No continuum emission was detected with a 5$\sigma$ upper limit 
of 0.8 mJy from TH2.

\begin{figure}
\begin{center}
\resizebox{\hsize}{!}{\includegraphics[bbllx=0cm,bburx=14.0cm,bblly=0cm,bbury=28.2cm,clip=,angle=0]{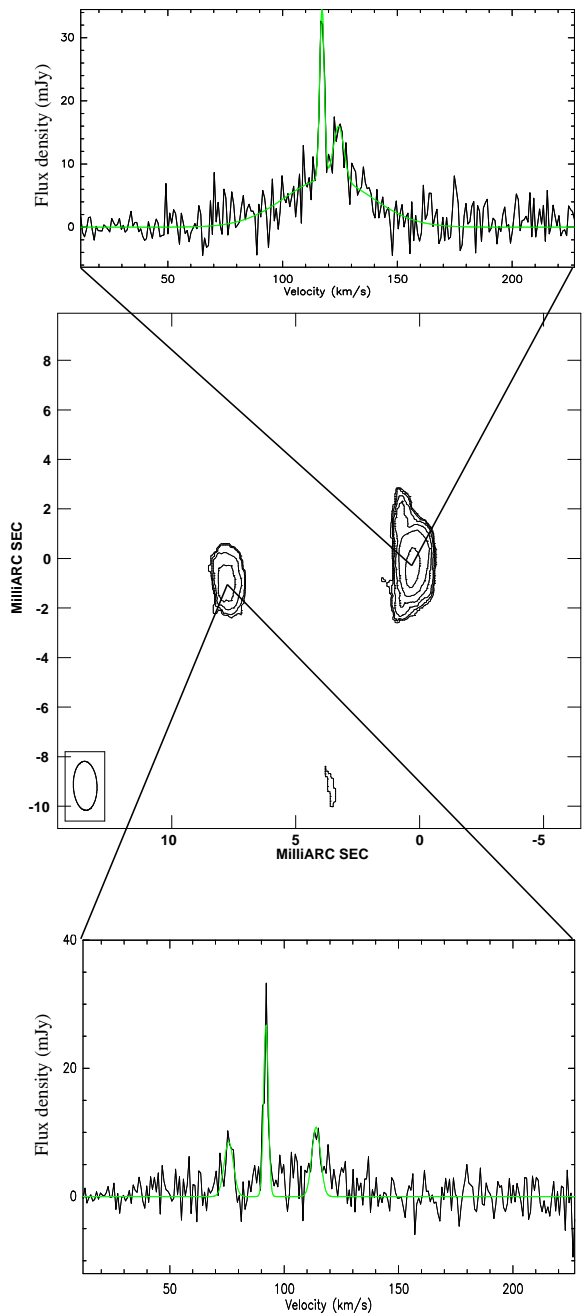}}
\end{center}
\caption{{\bf Middle Panel:} Velocity integrated water maser emission in NGC\,253 from the VLBI observation on 2008 March 24 at the position of TH4. The contours start at 7 mJy \kms\, and increase with factors of 2. {\bf Upper Panel:} Spectrum of the western feature, containing one broad and two narrow components. {\bf Lower Panel:} Spectrum of the eastern feature, containing three narrow components.}
\label{mom0}
\end{figure}

The frequency band with the maser emission was imaged channel by channel 
in a field of 0.2$\times$0.2 arcsecond$^2$ centered at the position of TH4.
The restoring beam was 1.98$\times$0.95 mas$^2$ with a position angle of 
2.3$^\circ$. We reached a noise level of $\sim$2.5 mJy in a single spectral
channel. We detected maser emission from two spatially distinct regions 
separated by 7.4 mas or 0.12 pc (VLBI-East and VLBI-West). A velocity 
integrated maser map is shown in Fig.\ref{mom0}. 

The spectrum of the western feature (Fig.~\ref{mom0}, upper panel) shows two 
narrow components (2--5 \kms) superimposed on a broad $\sim$45 \kms\ wide 
component.  The spectrum of the eastern feature (Fig.~\ref{mom0}, lower panel) 
shows three narrow (2--4 \kms) components, one with a flux density of 
$\sim$ 30 mJy and two with flux densities of $\sim$ 10 mJy. The parameters 
of Gaussian fits to the lines are summarized in Table~\ref{spec-data}. Please
note that the width of the line at v$_{\mathrm LSR}$=114 \kms\ in the eastern 
component was fixed in the fit, since otherwise the component becomes very 
broad (33 \kms, probably caused by the slightly higher noise in the 
spectrum of the eastern component at the velocity of the peak of the western 
component at 117 \kms). The positions of the emission from the three 
components in each spectrum are consistent within 0.2\,mas, the error being 
dominated by the position uncertainties of the weaker features.

The estimated absolute position of the strongest maser in the western 
feature is  00$^{\mathrm h}$47$^{\mathrm m}$33.16974$^{\mathrm s}$  
$-25^\circ$17'17.4734'' (J2000) with an uncertainty (the error is dominated 
by the uncertainty in the position reference) of 0.23 and 0.49 mas in 
right ascension and declination, respectively. This is consistent 
within 30 mas (or $\sim$0.5 pc) with the position of the continuum 
source TH4 from the VLA observations (see section~\ref{ac753}). This confirms
the result of HBT06 that the maser is associated with the continuum source 
TH4.

\begin{table*}
\caption{Details of the detected water maser components in our VLBI and VLA A-Array observations.}
\label{spec-data}
\centering
\begin{tabular}{lccccc}
\hline\hline
Source& v$_{\mathrm LSR}$& Width&  Peak Flux& Integrated Flux& L$_{\mathrm H_2O}$\\
      & [\kms] & [\kms]    & [mJy]    & [Jy \kms]& [L$_\odot$]\\
\hline
\\
VLBI-West & 117.1$\pm$0.1&  2.3$\pm$0.2& 27.4 &  0.068$\pm$0.007 &0.017\\
  & 119.9$\pm$1.5 & 44.6$\pm$6.3 &  7.7 & 0.366$\pm$0.026 & 0.092\\  
  & 124.4$\pm$0.4 &  5.3$\pm$1.3 &  8.5 &  0.048$\pm$0.014 & 0.012\\
\hline
\\
VLBI-East &  75.7$\pm$0.6&  4.8$\pm$1.5&  8.5 &  0.043$\pm$0.009&0.011 \\
  &  92.0$\pm$0.1 &  2.4$\pm$0.3 & 27.2 &   0.070$\pm$0.006 & 0.017 \\ 
  & 114.0$\pm$0.4 &  4.5         & 10.8 &   0.052$\pm$0.006 & 0.013 \\
\hline
\\
VLBI total&          &              &      &    0.647$\pm$0.033   & 0.162 \\
\hline
\\
VLA  & 116.8$\pm$0.5&   35$\pm$1 & & 2.10$\pm$0.06 &0.53\\
\hline\hline
\end{tabular}
\end{table*}

\subsection{VLA Data}
\subsubsection{2004 November 2 (A configuration)}
\label{ac753}
\begin{figure}
\begin{center}
\resizebox{\hsize}{!}{\includegraphics[angle=0]{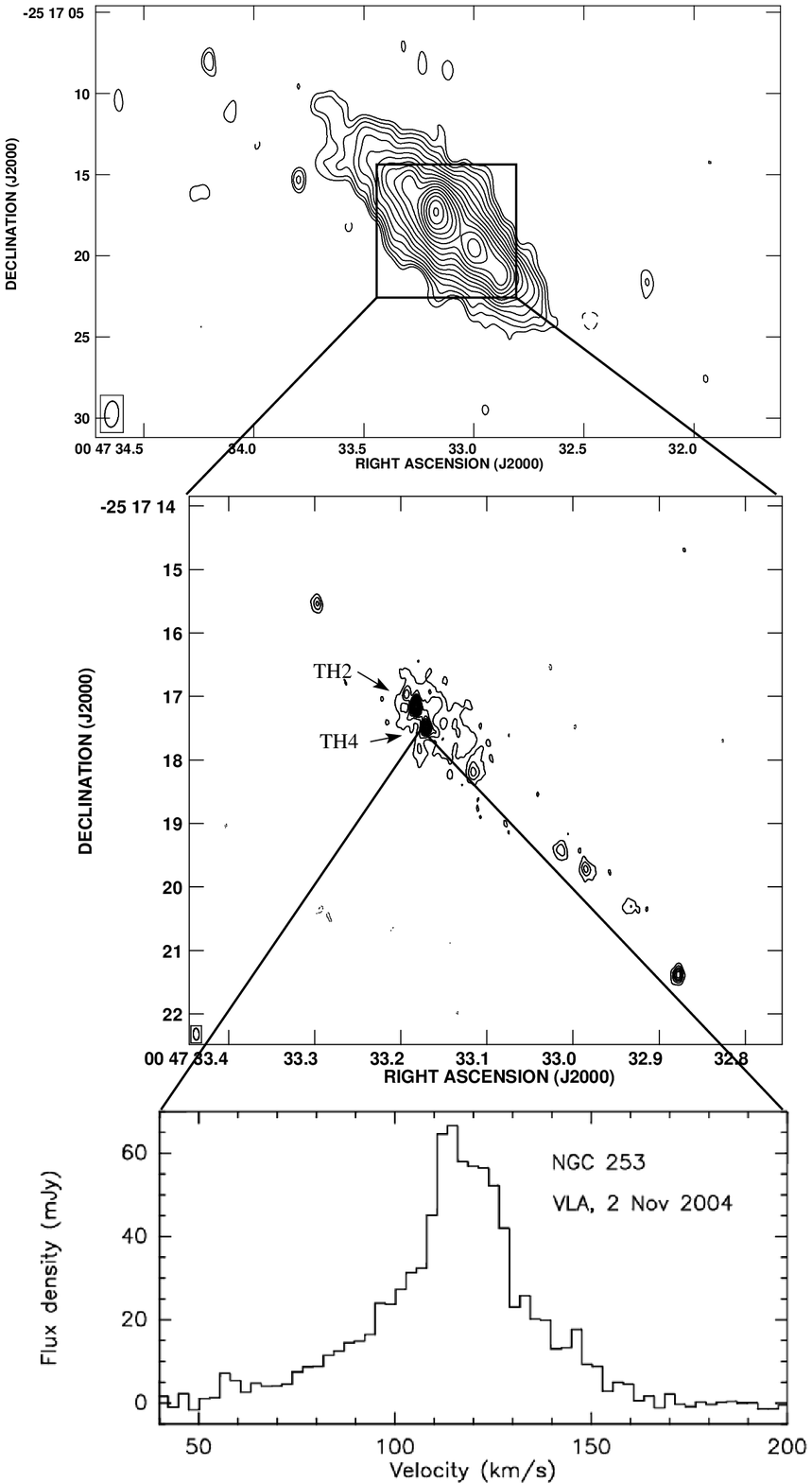}}
\end{center}
\caption{{\bf Upper Panel:} C-array VLA image of NGC\,253 at 22 GHz observed 
on 2008 March 24. The contours start at 0.4 mJy and increase by factors of 
$\sqrt{2}$. {\bf Middle Panel:} VLA A-array map of the continuum emission at 
22\,GHz of NGC\,253 observed on 2004 November 2. The synthesized beam is shown 
in lower left corner. Contours are at intervals of 0.45 mJy/beam, beginning 
at 1.35 mJy/beam (3$\sigma$).  
{\bf Lower panel}: VLA A-Array spectrum of the water maser detected towards NGC\,253 observed on 2004 November 2. The channel spacing is $\sim$\,2.6\,{\kms}.}
\label{fig:n253_vla}
\end{figure}

We detected water maser emission in NGC\,253 at $V_{\rm LSR} \sim$\,120\,{\kms}
with a peak flux density of $\sim$\,70\,mJy (Figure~\ref{fig:n253_vla}, lower 
panel). The maser line is confidently identified with the {\water} component 
reported by \citeN{HenkelTarchiMenten2004}. The line parameters, obtained from 
a Gaussian fit, 
are reported in Table~\ref{spec-data}. The line integrated intensity of 
$\sim$\,2.1\,Jy\,{\kms} yields an {\water} isotropic luminosity of 
$\sim$\,0.53\,L$_\odot$.

We produced a continuum map using the line free channels 
(Fig.~\ref{fig:n253_vla}). The synthesized beam was 0.18$\times$0.09 
arcseconds$^2$  with a position angle of 3$^\circ$. We detected 9 compact 
radio sources most of which can be identified with the brightest compact 
components reported in the nuclear region of NGC\,253 by UA97. One further 
source is present between TH8 and TH9 that was also detected by UA97 but 
had never assigned a label likely because of its weakness. Among the bright 
UA97 sources, only TH5 is not seen in our map.  
We fitted elliptical Gaussians to the source brightness distributions and 
derived the peak brightnesses and integrated flux densities as well as the 
angular dimensions. The results of these fits are reported in 
Table~\ref{VLA-sources}.

\subsubsection{2008 March 24 (C configuration)}
From the continuum data, we imaged a field of 102$\times$102 arcseconds$^2$, 
covering almost the full primary beam of a single VLA antenna. The restoring 
beam was 1.60$\times$0.85 arcseconds$^2$ with a position angle of $-$5$^\circ$.
The 22 GHz continuum emission of NGC\,253 is shown in Fig.~\ref{fig:n253_vla}. 
We achieved an image rms of 0.11 mJy\,beam$^{-1}$ which is the deepest 
continuum image of NGC\,253 at 22 GHz so far.  We also detected 
3 compact sources  that were not seen in the higher resolution maps of UA97
at 22 GHz: 4.81--43.60, 6.40--37.1, and 6.81--29.6 (using nomenclature of UA97,
 see Table~\ref{VLA-sources}).

From the line data, we imaged also a field of 102$\times$102 arcseconds$^2$
and achieved an image rms of 12 mJy. The restoring beam was 1.52$\times$1.30 
arcseconds$^2$ with a position angle of $-$11.61$^\circ$. No maser emission
was detected at a 5$\sigma$ upper limit of 60 mJy, indicating a lower flux
level of the maser compared to previous observations.

\begin{table*}
\caption{Details of the detected compact VLA sources in our VLA observations on 2004 November 2 (A configuration) and 2008 March 24 (C configuration). Shown are also the flux densities at 22 GHz from UA97 from their observations with the A and B configuration.}
\label{VLA-sources}
\centering
\begin{tabular}{lcccccccccc}
\hline\hline
Config&Source&$\alpha_{J2000}$ & $\delta_{J2000}$& Major axis& Minor axis& P.A. & Peak Flux& Integrated Flux& S$_\nu$ (VLA-A) &S$_\nu$ (VLA-B)\\
& &  00$^{\rm h}$ 47$^{\rm m}$ & -25\degr17\arcmin & [\arcsec] & [\arcsec] & [$^\circ$]& [mJy]    & [mJy] &[mJy]&[mJy]\\
\hline
\\
A & TH1        & 33\fs30 & 15\farcs54 & 0.24$\pm$0.02  & 0.17$\pm$0.02  &  16$\pm$10 &  3.8$\pm$0.4 &  9.1$\pm$1.2 &  2.09 &  6.72\\
A & TH2        & 33\fs18 & 17\farcs15 & 0.22$\pm$0.004 & 0.13$\pm$0.002 &   4$\pm$1  & 23.6$\pm$0.4 & 38.4$\pm$0.9 & 21.53 & 35.79\\
A & TH4        & 33\fs17 & 17\farcs48 & 0.26$\pm$0.007 & 0.16$\pm$0.005 &   6$\pm$2  & 12.5$\pm$0.4 & 29.9$\pm$1.2 &  3.78 &  9.69\\
A & TH6        & 33\fs11 & 18\farcs16 & 0.40$\pm$0.04  & 0.24$\pm$0.02  & 176$\pm$7  &  3.9$\pm$0.4 & 21.9$\pm$2.3 &  0.94 &  7.88\\
A & TH7        & 33\fs01 & 19\farcs42 & 0.26$\pm$0.03  & 0.19$\pm$0.02  &  15$\pm$13 &  3.7$\pm$0.4 & 10.7$\pm$1.4 &  0.88 &  5.84\\
A & TH8        & 32\fs98 & 19\farcs73 & 0.28$\pm$0.03  & 0.20$\pm$0.02  &  21$\pm$11 &  3.9$\pm$0.4 & 13.0$\pm$1.5 &  1.82 &  5.72\\
A &            & 32\fs93 & 20\farcs31 & 0.26$\pm$0.04  & 0.25$\pm$0.04  &  48$\pm$90 &  2.4$\pm$0.4 &  9.2$\pm$1.7 & --    & --   \\
A & TH9        & 32\fs88 & 21\farcs38 & 0.21$\pm$0.01  & 0.13$\pm$0.006 & 177$\pm$4  &  7.8$\pm$0.4 & 13.1$\pm$0.9 &  3.57 &  9.79\\
C & 4.81--43.6 & 32\fs22 & 21\farcs7  & 3.5$\pm$0.7    & 1.1$\pm$0.2    &   1$\pm$5  &  0.5$\pm$0.1 &  1.4$\pm$0.4 & --    & --   \\
C & 6.40--37.1 & 33\fs79 & 15\farcs3  & 1.5$\pm$0.2    & 0.9$\pm$0.1    & 177$\pm$9  &  0.9$\pm$0.1 &  0.8$\pm$0.2 & --    & --   \\
C & 6.81--29.6 & 34\fs20 & 08\farcs1  & 1.9$\pm$0.3    & 0.8$\pm$0.1    &   8$\pm$6  &  0.8$\pm$0.1 &  0.9$\pm$0.2 & --    & --   \\
\\
\hline\hline
\end{tabular}
\end{table*}

\section{Discussion}
\subsection{Continuum emission}

The assumed nucleus of NGC\,253 (TH2) has a brightness temperature grater than
20,000 K at 22 GHz and greater than 40,000 K at 15 GHz (UA97). Chandra 
observations of the nuclear region show evidence for an intermediate mass 
black hole or a low-luminosity AGN (LLAGN) component, but the spatial 
resolution is not sufficient to separate TH4 and TH2 
\cite{WeaverHeckmanStrickland2002}. The X-ray emission is also consistent 
with emission from ultra-luminous X-ray sources in other galaxies. Hence, 
it is still not clear whether the radio source TH2 is indeed an LLAGN.
Recently, Fern\'andez-Ontiveros et al. \citeyear{FernandezOntiveros2009} 
questioned the presence of an AGN component because of the lack of any optical 
or IR counterpart.

Variability of TH2 or TH4 would point toward an AGN nature of these radio 
sources. A comparison of our measured flux densities with values in the 
literature is difficult due to potential contamination by the extended diffuse 
emission from the galaxy. Different 
resolutions will lead to different flux density estimates for the same source. 
This can be seen in the comparison of the flux densities from A and B 
configuration data in UA97. The beam size in our observation is $\sim$40\% 
larger than the beam size in the A configuration observation of UA97 which 
explains why we find larger fluxes for all sources. Our flux density 
of 23.6 mJy for TH2 is consistent with the 21.5 mJy found in UA97, indicating
 no strong variability. However, HBT06 report a flux density of 
only $\sim$8 mJy for TH2 (VLA A configuration, and a synthesized beam 
comparable to our A configuration beam). On the other hand, our flux 
density for TH4 of 12.5 mJy is comparable to the values found by HBT06 
(8 mJy), but larger than the value reported in UA97 (3.78 mJy 
in A configuration and 9.69 mJy in B configuration). Thus, taking into account 
the different resolutions in the observations, there is no clear evidence
for variability of TH4 but signs of possible variability of TH2. Multi-epoch
observations with the VLA with identical resolutions are needed to decide 
this question.

The continuum source TH2, the presumed nucleus, is unresolved with the VLA 
with a source size $<100$ mas and a flux density of $\sim$20 mJy, but 
completely resolved by our VLBI observation. \citeN{Tingay2004} did not 
detect TH2 with the Australian Long Baseline Array (LBA) at 1.4 GHz 
probably due to strong free-free absorption. More recently, 
\citeN{LencTingay2006} performed higher frequency LBA observations at 2.3 GHz, 
detecting six sources. TH2 was detected with a flux density of $\sim$ 6 mJy, a 
source size of 80$\times$60 mas, and a brightness temperature 
of $\sim$ 10$^5$ K. This value is not particularly unusual and can be justified
as well in terms of radio supernova or a young supernova remnant without 
necessarily invoking the presence of AGN activity. The spectrum of the source 
can be fit with a free-free absorbed-power-law.

One possible reason for our non detection of TH2 could be scatter broadening 
similar to Sgr~A* (e.g. 
\citeNP{KrichbaumZensusWitzel1993,Yusef-ZadehCottonWardle1994}).
Assuming a $\lambda^2$ law and a source size of 80$\times$60 mas at 
2.3 GHz \cite{LencTingay2006}, one would expect a source size of 
$\sim 0.86\times0.64$ mas at 22 GHz, which is smaller than our beam 
size. Hence, we can rule out scatter broadening.

Since TH2 has a turnover frequency between 4.8 and 8 GHz (UA97), it could be 
a GHz peaked spectrum (GPS) source, i.e. a very young AGN . However, the 
linear size vs. turnover frequency relation \cite{ODeaBaum1997} predicts a 
size of $\sim$20 pc, more than 12 times larger than the upper limit of 1.6 pc
from the VLA observations. Nevertheless, there is considerable scatter in the 
correlation and at least one GPS source is known with a turnover frequency of 
10 GHz and a projected size of $<$ 1 pc 
\cite{BrunthalerFalckeBower2000,BrunthalerFalckeBower2005}. Hence, we can not
rule out a GPS scenario for NGC\,253, although GPS sources are usually several 
orders of magnitudes more powerful than TH2. The existence of an inactive 
supermassive black hole like Sgr A* can not be ruled out by our observation, 
since it's flux density at a distance of 3.3 Mpc would be only 6 $\mu$Jy.

\citeN{LencTingay2006} did not detect the source TH4, the continuum source
associated with the water maser emission,  at 2.3 GHz with the LBA, excluding 
with high confidence that TH4 is an AGN. UA97 suggest that TH4 is a SNR 
mostly because its radio spectral index ($\alpha_{1.3 \rm cm}^{3.6 \rm cm} 
= -0.52\pm0.17$) is typical for optically thin synchrotron emission.

\subsection{H$_2$O maser emission}
The upper limit of 60 mJy for maser emission in our VLA data that was taken 
during our VLBI observation in March 2008 shows that the maser (which we 
detected at about 30 mJy) was observed during a weak period. The spectrum of
 the eastern component in our 
VLBI observation resembles remarkably well the spectrum of the Galactic water 
maser in IC\,1396N \cite{SlyshValttsMigenes1999} where the blue-shifted and 
redshifted lines are found 410 and 10$^4$ AU from the position of the protostar
and are interpreted as an outflow. If one would interpret the 
spectrum as a rotating disk, a rotation velocity of $\sim$19 \kms\ 
($\frac{114-76}{2}$ \kms) at a radius of $\le$ 330 AU yields an 
enclosed mass of $\le$ 270 sin$^{-2}$~i~M$_\odot$. Altough the current data 
is not sufficient to distinguish between these two scenarios, the emission
is comparable with water masers in Galactic star forming regions. This is in
contrast to the broad emission ($\sim$ 45 \kms) in the western maser 
component, which is not typical for water masers in star forming regions. 
However, this broad component could be also the sum of several narrow 
components which blend together.

A comparison between the VLA A-array maser feature, observed by us in 
November 2004, with the one detected by \citeN{HenkelTarchiMenten2004} using 
the CnB configuration in September 2002 shows that both the line profile and 
the center velocity of the maser features are consistent. However, the 
integrated flux density and, as a consequence, the total isotropic luminosity 
of the line in 2004 are weaker by a factor of two than in 2002. Nevertheless, 
the line in our observation (see Table~\ref{spec-data}) is stronger than 
the line seen in 1997 (1.65\,Jy\,{\kms}, VLA CnB configuration) and 1998 
(0.54\,Jy\,{\kms},VLA A configuration) by HBT06. Thus, the maser exhibits 
significant variability on timescales of years.

\subsection{Association of the maser emission}

The water maser lines are blue shifted  on the order of 100--200 \kms\ with 
respect to the velocity of the gas in the masing region. This has been 
interpreted according to different scenarios that mostly depend on the 
presence  or absence of AGN activity in the nuclear region of NGC\,253.
The discrepancy in velocity might be explained by the presence of a 
circumnuclear accretion disk or an AGN-related nuclear outflow. However, 
as described in section 4.1, it is questionable whether NGC\,253 hosts an 
AGN. Since the water maser is clearly  associated with TH4 (HBT06; present 
work), this suggests that a maser-AGN relation is highly unlikely and that the 
maser emission is almost surely related to star formation activity.

\subsubsection{Association with the supernova remnant TH4?}
Given the spatial coincidence of the maser emission with TH4 (0.5 pc), 
which shows characteristics compatible with those of an SNR (Section 4.1), 
an association of the masing gas with the expanding shell of 
a SNR may seem promising. The anomalous velocity of the maser 
line would be due to the motion of the approaching side of the shell. 
However, the possibility that the maser is excited by the SNR itself is 
unlikely, since Galactic water masers have never been detected towards SNRs. 
A search for water masers in the three SNRs, W28, W44, and IC\,443, yielded 
no detections \cite{ClaussenGossFrail1999}. The authors explained the non 
detections as a consequence of low post-shock densities found in the shocked 
regions of the SNRs, which are too low to produce a strong enough inversion 
of the 22\,GHz maser line. However, the SNR could play an indirect role in 
the exciting mechanism. The young stellar object responsible for the excitation
of the maser might be born in a star forming region induced by the supernova 
shock wave like the scenario proposed for the Galactic {\it cometary} H{\sc ii}
region G34.3+0.2 \cite{ReidHo1985} which also host water masers (e.g. 
\citeNP{FeyGaumeNedoluha1994}) and is believed to interact with a wind from 
the precursor of the supernova remnant W44. 
Alternatively, the masing gas can be associated only in projection with the 
SNR. Here the  line can result from amplification of the radio 
continuum background from the SNR as suggested by HBT06. However, in this case
the masing gas should be still very close to the SNR shell to experience the 
`kick' or the line velocity issue would still remain open.

\subsubsection{Association with peculiar gas components?}
Other signs of peculiar kinematic gas components in NGC\,253 have 
been reported in a number of recent papers.  A bar-like structure, covering 
the inner 150$''$ (2.4 kpc) region of NGC\,253 was first observed in the
near-infrared 
(\citeNP{ScovilleSoiferNeugebauer1985}; \citeNP{ForbesDepoy1992}) and   
confirmed by subsequent optical and mid-infrared observations 
(\citeNP{ForbesDepoy1992}; \citeNP{PinaJonesPuetter1992}). Counterparts of 
this stellar bar have been also found in CO \cite{CanzianMundyScoville1988}, 
HCN \cite{PaglioneTosakiJackson1995}, and CS \cite{PengZhouWhiteoak1996}. 
Observations in the radio recombination lines (RRL) H92$\rm \alpha$ and 
H53$\rm \alpha$ at higher spatial resolution revealed a velocity field in 
the central 10$''$ (160 pc) that is discrepant with that derived from the
lower resolution CO, CS, and HCN observations (\citeNP{AnantharamaiahGoss1996};
\citeNP{RodriguezRicoGossZhao2006}). 

To explain the 
different kinematics observed for the ionized (traced by the RRLs) and the 
molecular gas (CO, HCN, and CS), three alternative scenarios have been
proposed including the presence of a secondary bar within the primary one, 
a compact supermassive object in the nuclear region of the galaxy, or 
the movement of the ionized gas in a starburst-driven outflow.
The first possibility is discussed in \citeN{RodriguezRicoGossZhao2006}, 
with the conclusion that it is 
unlikely  (although not a priori rejectable) because of the lack of the 
characteristic S-shape in the velocity field for the secondary bar. 
Because the lack of an AGN does not imply the absence of a supermassive 
object, the second option can not be ruled out by our conclusions that 
the maser is not related to an AGN.

The third option, an anomalous structure associated with outflow material 
seems viable. Starburst-driven outflows are indeed present in (starburst)
galaxies like NGC\,253, and they are produced by the mechanical energy 
released in the ISM from strong stellar winds and supernova explosions 
(e.g. \citeNP{NakaiHayashiHanda1987} for M\,82; 
\citeNP{GreveNeiningerTarchi2000} for NGC\,2146). In addition, since the main 
excitation mechanism for water masers has been singled out to be collisional 
pumping generated in shocks (e.g. \citeNP{Lo2005}), the maser in NGC\,253 
could be related to the 
outflow itself, sharing with its ionized gas component the blueshifted 
velocity. A further issue in favor of the outflow option is that the majority 
of kilomasers are indeed related to particularly intense star formation 
activity that is also the necessary ingredient for outflow phenomena. Hence, 
it is not surprising that the maser line velocity is more 
consistent with that of the ionized gas than with that of the other molecular 
tracers not participating in the outflows. 

As a last point on the relation between the kinematics of different gas 
components and maser origin, it is worth mentioning that large cavities, 
like bubbles and superbubbles, are produced in galactic disks by powerful 
starburst events that are detectable both in the atomic (HI) and 
molecular (CO) gas distribution (e.g. \citeNP{WeissWalterNeininger1999} 
for the case of M82). Very recently, \citeN{SakamotoHoIono2006} detected two 
shells or cavities interpreted as superbubbles (labeled SB1 and SB2) in 
the CO nuclear disk of 
NGC\,253, with diameters of $\sim$ 100 pc each and velocity structures 
indicative of expansion velocities of $\sim$ 50 \kms. The center of the
superbubble SB1 
is located South-West with respect to source TH4 and the masers.
Although the projected distance is too large to relate it to the maser, the 
contribution of similar structures should be taken into account when 
considering the velocity field anomalies and, in any case, 
\citeN{SakamotoHoIono2006} emphasize the presence of disturbances and the 
strong kinematic impact of the nuclear starburst on the central molecular 
disk hosting the starburst. This can, by itself, qualitatively motivate 
the anomalous velocities in the masing gas.


\section{Summary}
We have performed radio interferometic observations at 22 GHz with the VLBA and
the VLA of continuum sources and H$_2$O masers in the nucleus of the galaxy 
NGC\,253. The maser emission was found in two very compact ($<330$ 
AU) regions separated by 0.12 pc and is almost surely related to star formation
activity.

The continuum sources, while unresolved with the VLA, were not detected on mas
scales with an upper limit of 1 mJy (5$\sigma$). This questions the presence 
of a possible low-luminosity active galactic nucleus which has been previously
assumed to explain the radio emission from the brightest radio source in the 
central part of NGC\,253. Thus, the central region of NGC\,253 is 
dominated by a starburst, and not by an AGN.


\bibliography{brunthal_refs}
\bibliographystyle{aa}

\end{document}